\documentclass[conference]{IEEEtran}
\IEEEoverridecommandlockouts
\usepackage{cite}
\usepackage{amsmath,amssymb,amsfonts}
\usepackage{algorithmic}
\usepackage{graphicx}
\usepackage{textcomp}
\usepackage[nottoc]{tocbibind}
\usepackage{multirow}
\usepackage{xcolor}
\usepackage{booktabs}
\usepackage{threeparttable}
\usepackage{graphicx}
\usepackage{subfigure}
\def\BibTeX{{\rm B\kern-.05em{\sc i\kern-.025em b}\kern-.08em
    T\kern-.1667em\lower.7ex\hbox{E}\kern-.125emX}}
\begin{document}

\title{Binary Single-dimensional Convolutional Neural Network for Seizure Prediction\\
}

\author{\IEEEauthorblockN{Shiqi Zhao$^{1,2}$, Jie Yang$^{2}$, Yankun Xu$^{2}$, and Mohamad Sawan$^{2}$, \IEEEmembership{Fellow, IEEE}}
\IEEEauthorblockA{$^{1}$Zhejiang University, Hangzhou, Zhejiang, China 310058 \\
$^{2}$CenBRAIN Lab., School of Engineering, Westlake University, Hangzhou, Zhejiang, China 310024 \\
\ Email: zhaoshiqi@westlake.edu.cn}
}

\maketitle

\begin{abstract}

Nowadays, several deep learning methods are proposed to tackle the challenge of epileptic seizure prediction. However, these methods still cannot be implemented as part of implantable or efficient wearable devices due to their large hardware and corresponding high-power consumption. They usually require complex feature extraction process, large memory for storing high precision parameters and complex arithmetic computation, which greatly increases required hardware resources. Moreover, available yield poor prediction performance, because they adopt network architecture directly from image recognition applications fails to accurately consider the characteristics of EEG signals. We propose in this paper a hardware-friendly network called Binary Single-dimensional Convolutional Neural Network (BSDCNN) intended for epileptic seizure prediction. BSDCNN utilizes 1D convolutional kernels to improve prediction performance. All parameters are binarized to reduce the required computation and storage, except the first layer. Overall area under curve, sensitivity, and false prediction rate reaches 0.915, 89.26\%, 0.117/h and 0.970, 94.69\%, 0.095/h on American Epilepsy Society Seizure Prediction Challenge (AES) dataset and the CHB-MIT one respectively. The proposed architecture outperforms recent works while offering 7.2 and 25.5 times reductions on the size of parameter and computation, respectively.

\end{abstract}


\section{Introduction}
Up to 35\% of around 60 million epileptic patients are not under effective medical treatment due to the drug refractory\cite{ngugi2010estimation}, \cite{kwan2000early}, \cite{assi2017towards}. Epileptic patient may suffer from severe comorbidities, injuries and anxiety due to sudden seizure onset\cite{racine1972modification}. Hence, it is important to have an effective method of seizure prediction. EEG signals, commonly used for seizure prediction, can represent brain activity of epileptic patient\cite{mirowski2009classification}. The recorded typical EEG signals of an epileptic patient can be divided into four states: Interictal (between seizures), Preictal (before seizure), Ictal (seizure) and Post-ictal (after seizure)\cite{mormann2006seizure}. The preliminary goal of seizure prediction is to distinguish between interictal and preictal states. 
Most recently published seizure prediction methods are based on EEG or Intracortical EEG signals include two main steps. The first one is called feature extraction, which is used to extract features from the raw signals\cite{shahnaz2015seizure}.
For example, short-time Fourier transform (STFT) is used to transfer time domain signals to frequency domain features\cite{truong2018convolutional}. The second step consists of either classification based on the selected features, algorithms such as Rule-based decision\cite{assi2017towards}, Threshold Crossing\cite{eftekhar2014ngram} and Support Vector Machine (SVM)\cite{sharif2017prediction}.
 
Recently, deep learning algorithms have been used for EEG signals analysis, where the most representative algorithm is Convolutional neural network (CNN). Truong et al. used STFT and CNN with 2D convolution to process both EEG and IcEEG signals\cite{truong2018convolutional}. Eberlein et al. processed raw EEG signals time domain with a deep CNN to predict seizure onset\cite{eberlein2018convolutional}.  Truong et al. used Integer CNN and binary weights CNN for seizure detection, where state-of-the-art results were achieved\cite{truong2018integer}. Hossain et al. used 1D and 2D mixed convolution for seizure detection and got good results\cite{hossain2019applying}.

Although some good results have been obtained, these algorithms still have some drawbacks. Firstly, many deep learning classification algorithms still need extra feature extraction steps\cite{truong2018convolutional},\cite{tsou2019epilepsy}. Secondly, most reported works only adopt network architecture from computer vision and fail to consider the accurate characteristics of EEG signals\cite{korshunova2017towards}. It is important to notice that most algorithms for seizure prediction are not hardware-friendly oriented due to the large number of high precision floating point parameters and the corresponding complex computation\cite{marni2018real}.

In this paper, we propose Binary Single-dimensional Convolutional Neural Network (BSDCNN) trained with raw EEG data to predict seizure onset. Firstly, the conventional feature extraction is skipped in this work. Instead raw EEG data is directly used as input without any steps of preprocessing. Secondly, BSDCNN utilizes 1D convolution to better match the characteristic of EEG signals. Theoretical explanation is given. Thirdly, weights and activation values are binarized which reduce the scale of parameter and the computational complexity significantly. The remaining sections of this paper are organized as follows. Section II introduces proposed design method and used datasets. Results evaluation and comparison with other works are described in Section III. The last section concludes this paper.

\section{Proposed Method}
The purpose of seizure prediction is to distinguish between preictal and interictal brain states. Two popular datasets of seizure prediction were used in this research: the American Epilepsy Society Seizure Prediction Challenge (AES) \cite{brinkmann2016crowdsourcing} and the CHB-MIT one\cite{goldberger2000physiobank}. Details of the proposed BSDCNN algorithm will be described in this section.

\begin{figure*}[htbp]
\centerline{\includegraphics[width=7in,height=2in]{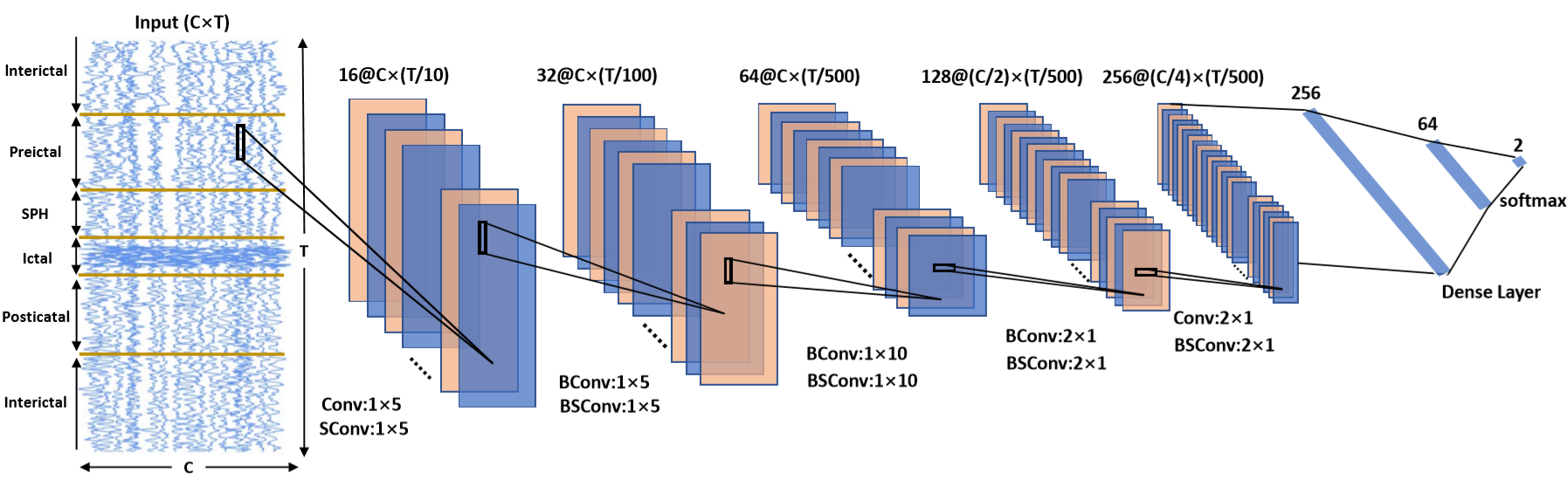}}
\caption{Structure of proposed model. Seizure prediction horizon (SPH) is a short interval between preictal interval and seizure onset. Except for intervals of preictal, ictal, SPH and postictal, the remaining parts of EEG signals belong to interictal interval. Interictal and preictal intervals are extracted for input of neural network.
There are five Convolution blocks, every block includes one convolution layer and one strided convolution layer with batch normalization. Except for parameters in the first convolution blocks are full precision and use ReLU activation function, all parameters in other blocks are binary and use signum activation function. Two fully connected layers with sigmoid activation function and one last fully connected layer with softmax function is used to get the output. When different subjects are used to input, the size of the features changes accordingly. }
\label{fig}
\end{figure*}

\subsection{Datasets}
AES dataset contains EEG data collected from five canines and two human subjects with epilepsy. The EEG data were recorded with 16 (or 15) electrodes and 400 (or 5000) Hz sampling rate\cite{karoly2017circadian}. As shown in Fig. 1, there are two important parameters, namely seizure prediction horizon (SPH) and preictal interval length (PIL)\cite{assi2015hybrid}. SPH defines the interval between preictal and seizure onset, while PIL is the length of preictal state. In this dataset, SPH and PIL are set to 5 minutes and 1 hour, respectively.
Samples are extracted from intericatal and preictal intervals respectively with fixed 20-seconds time window, then the ${16\ (or\ 15)}\times{8000}$ matrix is used as input data.

CHB-MIT dataset contains scalp EEG data of 23 measurements from 22 patients. All measurements are recorded at 256Hz sampling rate\cite{furbass2015prospective}. Fixed 23-electrode configurations are used in 15 measurements, while there are some changes in electrode configuration for the remaining measurements\cite{alickovic2018performance}. Moreover, we only consider measurements with no less than 3 lead seizures and exclude one subject (chb06) due to its absence from comparison with latest works. Consequently, only measurement from 6 subjects were selected in this work. We set 5min SPH and 30min PIL for fair result comparison to other state-of-the-art works. Fixed 20s time window is also used in this dataset, the shape of input data is ${23}\times{5120}$.

The ratio between preictal and interictal is about 5:1 and 4:1 in AES dataset and CHB-MIT dataset respectively. The training model is likely to be unsatisfactory if trained with imbalanced data. The solution to this challenge in this work is presented as follows. 

\subsection{Single-dimensional CNN Model}
Two-dimensional convolution kernel allows to deliver the excellent performance in image recognition. However, it hardly has requested best performance in seizure prediction for the following two reasons. Firstly, the two dimensions of EEG data are different from the image data. In image data, the two dimensions are both recording of pixels. However, the two dimensions of EEG data have different meanings of time and channel, respectively. We argue that mixing up the channel and time dimension will decrease the prediction performance. It is demonstrated in our experiment in section III.
Secondly, single-dimensional convolutional kernel has the same resolution for each line of input, while 2D convolutional kernel extracts less information of edge pixels compared with internal pixels. The edge sampling points of EEG signals contain equivalent information compared with interior sampling points, while the edge pixels in the image often contain less information than interior pixels. This is why 1D convolution kernel has great performance in image classification, but it has some drawbacks in EEG signals analysis.

\subsection{Binary Single-dimensional Convolutional Neural Network for Seizure Prediction}

Binary convolutional neural network (BCNN) uses binary activation values and weights in place of activation values and weights of full-precision. Generally speaking, the weights and activation values of BCNN are constrained to +1 or -1\cite{rastegari2016xnor}. In CNN, most of computation time and resource are intended for the Multiply Accumulate (MAC) operation, binarize the activation values and weights can reduce computation time and complexity significantly\cite{lin2017towards}. In addition, the hardware implementation of BCNN is easier than a full-precision CNN, it also has lower hardware resource and power consumption requirements\cite{yu2016binary}. 

In image classification task, BCNN has achieved good results when compared with full-precision networks. CIFAR10 $89.85\%$ (the full precision is $92.38\%$\cite{courbariaux2016binarized}) and YOLOv2 $67.6\%$ (the full precision is $69.1\%$\cite{nakahara2018lightweight}). Motivated by building a model with less parameter scale and less computation for seizure prediction, BCNN is used in this work.

Signum function is used for the binarization of weights and activations values. Because of the use of signum function, the gradient will be 0 when back propagation. To make the gradient can be propagated, we must limit the gradient by the following equation\cite{courbariaux2016binarized}:

\begin{equation}
Htanh(x) = Clip(x,-1,1) = max(-1,min(1,x)){}
\end{equation}

In back propagation, signum function is replaced by equation (1). It means that the gradient of this node is constrained to 1 when input is between -1 and 1, while the gradient of this node is 0 at other input values. It should be noted that in the training process, we also retain the full-precision weights. When back propagation, the full-precision weights are updated in back propagation, while the binarization are applied only in forward propagation. 
Since the impact on the difference in the distribution of data for each batch, the speed of convergence and training effect will be affected greatly when binarization of activation values. 
Batch normalization is added before signum activation function in every convolutional block, which improves the convergence speed and training effect of the model\cite{ioffe2015batch}.

Single-dimensional convolution will continue to be used in BCNN. However, we use smaller convolution kernels instead of previous ones and replace the pooling layer with strided convolution layer to get better performance. This unified form may have some advantages in hardware implementation. In addition to the first convolutional layer, strided convolution layer and dense layers, the parameters of the remaining layers are carried out to binarization. 

Based on the above methods, we propose binary single-dimensional convolutional neural network (BSDCNN), which combines the abilities of single-dimensional convolution for feature extraction of EEG signals and the advantages of computational complexity and power consumption of BCNN. The network structure is shown in Fig. 1. To reduce hardware resource consumption, the proposed network takes raw EEG signals as input. There are five convolution blocks. Each convolutional block consists of a convolution layer and a strided convolution layer. Batch normalization layer was added after every convolutional layer. Except for the first convolution block, the parameters of other convolution blocks are binary. 
The first three blocks are convolution of time dimension. The size of convolution kernels are ${1}\times{5}$, ${1}\times{5}$ and ${1}\times{10}$, respectively and the amount of convolution kernels are 16, 32 and 64, respectively. The last two convolutional blocks are convolution of channel dimension. The size of convolution kernels are ${2}\times{1}$ and the amount of convolution kernels are 128 and 256, respectively. Following that, there are three fully connected layers with sigmoid activation function and output sizes of 256, 64 and 2. The results and discussion are shown in section III.

\renewcommand{\arraystretch}{1.5} 
\begin{table}[htbp]
\caption{AUC comparisons in different modes of convolution }
\begin{center}
\begin{tabular}{c c c c c c}
\hline\hline
& \multirow{2}{*}{\textbf{1D--1D}} &  \multirow{2}{*}{\textbf{1D--2D}} & \multirow{2}{*}{\textbf{2D--1D}} &\multirow{2}{*}{\textbf{2D--2D}} \\
\\
\hline
Dog1  & 0.94 & 0.92 & 0.78 & 0.82 \\
Dog2  & 0.99 & 0.98 & 0.96 & 0.97 \\
Dog3  & 0.97 & 0.97 & 0.89 & 0.91 \\
\hline
Average &\textbf{0.977} & 0.957 & 0.877 & 0.900 \\
\hline\hline
\end{tabular}
\begin{tablenotes}
        \item[] 1D: Single-dimensional convolution; 2D: Two-dimensional convolution; 1D-2D: Single-dimensional convolution of time dimension and Two-dimensional convolution of channel dimension; AUC: Area Under Curve
\end{tablenotes}
\end{center}
\end{table} 

\subsection{Training}
In order to overcome problems of imbalanced dataset and improve the performance,  when we extract preictal samples to make up training dataset with 5s overlapping, but interictal samples are extracted without overlapping. This is equivalent to oversampling the small amount of preictal data.  In addition, the 16 training data are randomly extracted from both interictal and preictal for training. Accordingly, this method ensures that 32 training data of every batch will include the same number of two types of data. By using these methods, the influence of imbalanced data is reduced to some extent.

The model was trained for 20 epochs. For one epoch, the training process takes average 605s and 76s on AES and CHB-MIT dataset respectively. Proposed model is implemented in Python 3.6 with use of Keras 2.2 with a Tensorflow 1.13 backend on single NVIDIA 2080Ti GPU\cite{abadi2016tensorflow}.

\section{Results}
In order to compare with other state-of-the-art works, we evaluate our model with the following metrics: sensitivity, false prediction rate (FPR) and area under curve (AUC).

Table I shows the AUC results of using 1D or 2D convolution kernel in time or in channel dimensions. Previous researchers only mentioned that mixed convolution of time and spatial dimensions affect performance \cite{eberlein2018convolutional}, \cite{cecotti2010convolutional}.
 However, this does not apply when the used time dimension convolution kernel is based on two-dimensional structure. This is because the previous two-dimensional convolution in the first three convolutional blocks does not extract enough information for the first and last rows. Therefore, in the later convolution process, some useful information in time domain can still be extracted by using two-dimensional convolution kernel. The benefits of channel dimension one-dimensional convolution kernel are limited, because the number of channels is much smaller than the number of sampling points. If the single-dimensional time convolution kernel is used in first three convolutional blocks, then enough time information can be extracted, the use of single-dimensional channel convolution kernel in the following convolutional process will have a good performance.

\renewcommand{\arraystretch}{1.3} 
\begin{table}[tp]

  \centering
  \fontsize{6.5}{8}\selectfont
  \begin{threeparttable}
  \caption{Comparsion with other seizure prediction methods applied to AES Seizure Prediction Challenge dataset.}
  \label{tab:performance_comparison}
    \begin{tabular}{c c c c c c}
    \toprule
    \multicolumn{3}{c}{SDCNN}&\multicolumn{3}{c}{\bf BSDCNN}\cr
    \cmidrule(lr){1-3} \cmidrule(lr){4-6}
    Layer&Parameter type & Mem &Layer&Parameter type & Mem\cr
    \midrule
    Conv1 & Int & 10K & Conv1 & Int & 2.6K\cr
    Pool&--&--& SConv1 & Int & 40K\cr
    Conv2 & Int & 320K & BConv1 & Bin & 2.6K\cr
    Pool&--&--& BSConv1 & Bin & 5.1K\cr
    Conv3 & Int & 640K & BConv2 & Bin & 10.3K\cr
    Pool&--&--& BSConv2 & Bin & 20.3K\cr
    Conv4 & Int & 768K & BConv3 & Bin & 40.5K\cr
    Pool&--&--& BSConv3 & Bin & 80.5K\cr
    Conv5 & Int & 3.1M & BConv4 & Bin & 161K\cr
    Pool&--&--& BSConv4 & Bin & 321K\cr
    \hline
    Overall&--&4.84M&--&--&\textbf{672K}\cr
    \hline
    AUC&--&0.982&--&--&\textbf{0.955}\cr
    \bottomrule
    \end{tabular}
    \begin{tablenotes}
        \item[] Conv: Convolutional layer; Pool: Max-pooling layer; 
        
        AUC: Area Under Curve;
        SConv: Strided convolutional layer; 
        
        BConv: Binary Convolutional layer with batch normalization layer; 
        
        BSConv: Binary Strided Convolutional layer with batch normalization layer
    \end{tablenotes}
    \end{threeparttable}
\end{table}

\begin{figure}[htbp]
\centerline{\includegraphics[width=2.6in]{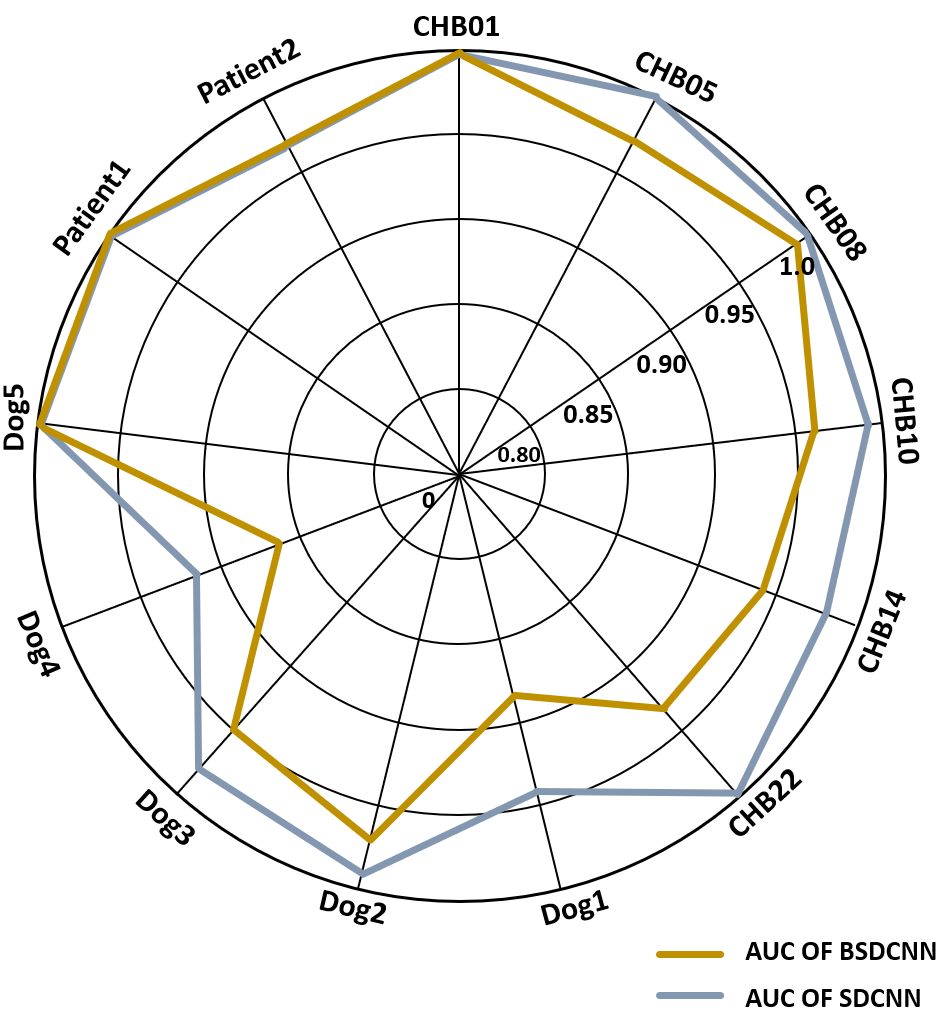}}
\caption{AUC comparison between SDCNN and BSDCNN based on different subjects. }
\label{fig}
\end{figure}

TABLE II compares the parameter scales of convolutional block between BSDCNN and high precision CNN with 32bit floating points called SDCNN. Compared to high precision CNN, 7.2 times reduction of parameter memory by using BSDCNN. Moreover, it also reduces the computation about 25.5 times through calculation of bit number multiplication and accumulation. It is important to note that all our calculations include batch normalization layer. The test results of SDCNN and BSDCNN in different datasets are shown in the Fig. 2 and the average AUC of SDCNN and BSDCNN are shown in TABLE IV. As can be seen from the results, using BSDCNN only reduces the average AUC of 2.74\% compared to using SDCNN.

\renewcommand{\arraystretch}{1.5} 
\begin{table}[tp]
  \centering
  \fontsize{6.4}{8}\selectfont
  \begin{threeparttable}
  \caption{Comparsion with other seizure prediction methods applied to AES Seizure Prediction Challenge dataset.}
  \label{tab:performance_comparison}
    \begin{tabular}{c c c c c c c c c}
    \toprule
    \multirow{2}{*}{Method}&
    \multicolumn{2}{c}{STFT+CNN\cite{truong2018convolutional}}&\multicolumn{2}{c}{ CNN\cite{eberlein2018convolutional}}&\multicolumn{3}{c}{\bf BSDCNN\ (This work)}\cr
    \cmidrule(lr){2-3} \cmidrule(lr){4-5} \cmidrule(lr){6-8}
    &FPR(/h)&SEN(\%)&AUC&SEN(\%)&AUC&SEN(\%)&FPR(/h)\cr
    \midrule
    Dog1&0.17&50&0.798&--&{\bf 0.88}&{\bf 77.90}&{\bf 0.23}\cr
    Dog2&0.01&100&0.812&--&{\bf 0.97}&{\bf 95.13}&{\bf 0.14}\cr
    Dog3&0.05&58.3&0.844&--&{\bf 0.95}&{\bf 81.64}&{\bf 0.07}\cr
    Dog4&0.41&78.6&0.919&--&{\bf 0.86}&{\bf 77.32}&{\bf 0.20}\cr
    Dog5&0.07&80&--&--&{\bf 1}&{\bf 98.41}&{\bf 0.03}\cr
    Pat1&0.36&100&--&--&{\bf 0.99}&{\bf 96.86}&{\bf 0.04}\cr
    Pat2&0.86&66.7&--&--&{\bf 0.97}&{\bf 97.58}&{\bf 0.11}\cr
    \hline
    Average&0.276&76.2&0.843&--&{\bf 0.915}&{\bf 89.26}&{\bf 0.117}\cr
    \bottomrule
    \end{tabular}
    \begin{tablenotes}
        \item[] STFT: Short-Time Fourier Transform; AUC: Area Under Curve; SEN:Sensitivity; FPR: False Prediction Rate
        
        Average AUC in this work is the mean AUC of Dog1 to Dog4. The average AUC of all subjects is 0.946. 
    \end{tablenotes}
    \end{threeparttable}
\end{table}

\renewcommand{\arraystretch}{1.5} 
\begin{table}[tp]
  \centering
  \fontsize{6.4}{8}\selectfont
  \begin{threeparttable}
  \caption{Comparsion with other seizure prediction methods applied to CHB-MIT dataset.}
  \label{tab:performance_comparison}
    \begin{tabular}{c c c c c c c c c}
    \toprule
    \multirow{2}{*}{Method}&
    \multicolumn{2}{c}{STFT+CNN\cite{truong2018convolutional}}&\multicolumn{2}{c}{ Wavelet+CNN\cite{khan2017focal}}&\multicolumn{3}{c}{\bf BSDCNN\ (This work)}\cr
    \cmidrule(lr){2-3} \cmidrule(lr){4-5} \cmidrule(lr){6-8}
    &FPR(/h)&SEN(\%)&AUC&SEN(\%)&AUC&SEN(\%)&FPR(/h)\cr
    \midrule
    chb01&0.24&85.7&0.943&--&{\bf 1}&{\bf 100}&{\bf 0.01}\cr
    chb05&0.16&80.0&0.988&--&{\bf 0.97}&{\bf 90.60}&{\bf 0.08}\cr
    chb08&--&--&0.921&--&{\bf 0.99}&{\bf 99.24}&{\bf 0.05}\cr
    chb10&0.00&33.3&0.855&--&{\bf 0.96}&{\bf 94.26}&{\bf 0.12}\cr
    chb14&0.40&80.3&--&--&{\bf 0.94}&{\bf 93.90}&{\bf 0.17}\cr
    chb22&--&--&0.877&--&{\bf 0.93}&{\bf 93.29}&{\bf 0.19}\cr
    \hline
    Average&0.20&69.83&0.917&--&{\bf 0.970}&{\bf 94.69}&{\bf 0.095}\cr
    \bottomrule
    \end{tabular}
    \begin{tablenotes}
        \item[] STFT: Short-Time Fourier Transform; AUC: Area Under Curve; SEN: Sensitivity; FPR: False Prediction Rate
        
        Average FPR, AUC and SEN in this work is the mean of corresponding samples. The average FPR, AUC and SEN of all subjects is 0.10, 0.965 and 95.22 respectively. 
    \end{tablenotes}
    \end{threeparttable}
\end{table}

Finally, we compare our model with other recent state-of-the-art works based on CNN. Table III and Table IV demonstrate comparisons of performance on AES and CHB-MIT datasets respectively. Truong et al. \cite{truong2018convolutional} and Khan et al.\cite{khan2017focal} use extra feature extraction steps, while Eberlein et al.\cite{eberlein2018convolutional} use raw EEG data directly. The proposed method achieved the highest AUC, sensitivity and FPR/h among others.

\section{Conclusion}
In this work, Binary Single-dimensional Convolutional Neural Network (BSDCNN) for seizure prediction has been proposed. By using binary neural network with single-dimensional convolutional kernels, the proposed model for seizure prediction model, compared with full-precision one, gets better results than all available models.
This method greatly reduces 7.2 times the size of the parameters and 25.5 times the computation complexity with only precision loss of 2.74\%. Comparing with recent state-of-the-art works, the proposed BSDCNN presents better performance. Moreover, the theoretical explanation of single-dimensional convolution shows better performance in seizure prediction compared with two-dimensional convolution model.

\section*{Acknowledgment}
The authors would like to acknowledge start-up funds from Westlake University to the Cutting-Edge Net of Biomedical Research and INnovation (CenBRAIN) to support this project.

{\small{
\bibliographystyle{IEEEtran}
\bibliography{reference.bib}}

\begin{thebibliography}{10}
\providecommand{\url}[1]{#1}
\csname url@samestyle\endcsname
\providecommand{\newblock}{\relax}
\providecommand{\bibinfo}[2]{#2}
\providecommand{\BIBentrySTDinterwordspacing}{\spaceskip=0pt\relax}
\providecommand{\BIBentryALTinterwordstretchfactor}{4}
\providecommand{\BIBentryALTinterwordspacing}{\spaceskip=\fontdimen2\font plus
\BIBentryALTinterwordstretchfactor\fontdimen3\font minus
  \fontdimen4\font\relax}
\providecommand{\BIBforeignlanguage}[2]{{%
\expandafter\ifx\csname l@#1\endcsname\relax
\typeout{** WARNING: IEEEtran.bst: No hyphenation pattern has been}%
\typeout{** loaded for the language `#1'. Using the pattern for}%
\typeout{** the default language instead.}%
\else
\language=\csname l@#1\endcsname
\fi
#2}}
\providecommand{\BIBdecl}{\relax}
\BIBdecl

\bibitem{ngugi2010estimation}
A.~K. Ngugi, C.~Bottomley, I.~Kleinschmidt, J.~W. Sander, and C.~R. Newton,
  ``Estimation of the burden of active and life-time epilepsy: a meta-analytic
  approach,'' \emph{Epilepsia}, vol.~51, no.~5, pp. 883--890, 2010.

\bibitem{kwan2000early}
P.~Kwan and M.~J. Brodie, ``Early identification of refractory epilepsy,''
  \emph{New England Journal of Medicine}, vol. 342, no.~5, pp. 314--319, 2000.

\bibitem{assi2017towards}
E.~Bou~Assi, D.~K. Nguyen, S.~Rihana, and M.~Sawan, ``Towards accurate
  prediction of epileptic seizures: A review,'' \emph{Biomedical Signal
  Processing and Control}, vol.~34, pp. 144--157, 2017.

\bibitem{racine1972modification}
R.~J. Racine, ``Modification of seizure activity by electrical stimulation: Ii.
  motor seizure,'' \emph{Electroencephalography and clinical neurophysiology},
  vol.~32, no.~3, pp. 281--294, 1972.

\bibitem{mirowski2009classification}
P.~Mirowski, D.~Madhavan, Y.~LeCun, and R.~Kuzniecky, ``Classification of
  patterns of eeg synchronization for seizure prediction,'' \emph{Clinical
  neurophysiology}, vol. 120, no.~11, pp. 1927--1940, 2009.

\bibitem{mormann2006seizure}
F.~Mormann, R.~G. Andrzejak, C.~E. Elger, and K.~Lehnertz, ``Seizure
  prediction: the long and winding road,'' \emph{Brain}, vol. 130, no.~2, pp.
  314--333, 2006.

\bibitem{shahnaz2015seizure}
C.~Shahnaz, R.~M. Rafi, S.~A. Fattah, W.-P. Zhu, and M.~O. Ahmad, ``Seizure
  detection exploiting emd-wavelet analysis of eeg signals,'' in \emph{2015
  IEEE International Symposium on Circuits and Systems (ISCAS)}.\hskip 1em plus
  0.5em minus 0.4em\relax IEEE, 2015, pp. 57--60.

\bibitem{truong2018convolutional}
N.~D. Truong, A.~D. Nguyen, L.~Kuhlmann, M.~R. Bonyadi, J.~Yang, S.~Ippolito,
  and O.~Kavehei, ``Convolutional neural networks for seizure prediction using
  intracranial and scalp electroencephalogram,'' \emph{Neural Networks}, vol.
  105, pp. 104--111, 2018.

\bibitem{eftekhar2014ngram}
A.~Eftekhar, W.~Juffali, J.~El-Imad, T.~G. Constandinou, and C.~Toumazou,
  ``Ngram-derived pattern recognition for the detection and prediction of
  epileptic seizures,'' \emph{PloS one}, vol.~9, no.~6, p. e96235, 2014.

\bibitem{sharif2017prediction}
B.~Sharif and A.~H. Jafari, ``Prediction of epileptic seizures from eeg using
  analysis of ictal rules on poincar{\'e} plane,'' \emph{Computer methods and
  programs in biomedicine}, vol. 145, pp. 11--22, 2017.

\bibitem{eberlein2018convolutional}
M.~Eberlein, R.~Hildebrand, R.~Tetzlaff, N.~Hoffmann, L.~Kuhlmann,
  B.~Brinkmann, and J.~M{\"u}ller, ``Convolutional neural networks for
  epileptic seizure prediction,'' in \emph{2018 IEEE International Conference
  on Bioinformatics and Biomedicine (BIBM)}.\hskip 1em plus 0.5em minus
  0.4em\relax IEEE, 2018, pp. 2577--2582.

\bibitem{truong2018integer}
N.~D. Truong, A.~D. Nguyen, L.~Kuhlmann, M.~R. Bonyadi, J.~Yang, S.~Ippolito,
  and O.~Kavehei, ``Integer convolutional neural network for seizure
  detection,'' \emph{IEEE Journal on Emerging and Selected Topics in Circuits
  and Systems}, vol.~8, no.~4, pp. 849--857, 2018.

\bibitem{hossain2019applying}
M.~S. Hossain, S.~U. Amin, M.~Alsulaiman, and G.~Muhammad, ``Applying deep
  learning for epilepsy seizure detection and brain mapping visualization,''
  \emph{ACM Transactions on Multimedia Computing, Communications, and
  Applications (TOMM)}, vol.~15, no.~1s, p.~10, 2019.

\bibitem{tsou2019epilepsy}
C.~Tsou, C.-C. Liao, and S.-Y. Lee, ``Epilepsy identification system with
  neural network hardware implementation,'' in \emph{2019 IEEE International
  Conference on Artificial Intelligence Circuits and Systems (AICAS)}.\hskip
  1em plus 0.5em minus 0.4em\relax IEEE, 2019, pp. 163--166.

\bibitem{korshunova2017towards}
I.~Korshunova, P.-J. Kindermans, J.~Degrave, T.~Verhoeven, B.~H. Brinkmann, and
  J.~Dambre, ``Towards improved design and evaluation of epileptic seizure
  predictors,'' \emph{IEEE Transactions on Biomedical Engineering}, vol.~65,
  no.~3, pp. 502--510, 2017.

\bibitem{marni2018real}
L.~Marni, M.~Hosseini, J.~Hopp, P.~Mohseni, and T.~Mohsenin, ``A real-time
  wearable fpga-based seizure detection processor using mcmc,'' in \emph{2018
  IEEE International Symposium on Circuits and Systems (ISCAS)}.\hskip 1em plus
  0.5em minus 0.4em\relax IEEE, 2018, pp. 1--4.

\bibitem{brinkmann2016crowdsourcing}
B.~H. Brinkmann, J.~Wagenaar, D.~Abbot, P.~Adkins, S.~C. Bosshard, M.~Chen,
  Q.~M. Tieng, J.~He, F.~Mu{\~n}oz-Almaraz, P.~Botella-Rocamora \emph{et~al.},
  ``Crowdsourcing reproducible seizure forecasting in human and canine
  epilepsy,'' \emph{Brain}, vol. 139, no.~6, pp. 1713--1722, 2016.

\bibitem{goldberger2000physiobank}
A.~L. Goldberger, L.~A. Amaral, L.~Glass, J.~M. Hausdorff, P.~C. Ivanov, R.~G.
  Mark, J.~E. Mietus, G.~B. Moody, C.-K. Peng, and H.~E. Stanley, ``Physiobank,
  physiotoolkit, and physionet: components of a new research resource for
  complex physiologic signals,'' \emph{Circulation}, vol. 101, no.~23, pp.
  e215--e220, 2000.

\bibitem{karoly2017circadian}
P.~J. Karoly, H.~Ung, D.~B. Grayden, L.~Kuhlmann, K.~Leyde, M.~J. Cook, and
  D.~R. Freestone, ``The circadian profile of epilepsy improves seizure
  forecasting,'' \emph{Brain}, vol. 140, no.~8, pp. 2169--2182, 2017.

\bibitem{assi2015hybrid}
E.~Bou~Assi, M.~Sawan, D.~Nguyen, and S.~Rihana, ``A hybrid mrmr-genetic based
  selection method for the prediction of epileptic seizures,'' in \emph{2015
  IEEE Biomedical Circuits and Systems Conference (BioCAS)}.\hskip 1em plus
  0.5em minus 0.4em\relax IEEE, 2015, pp. 1--4.

\bibitem{furbass2015prospective}
F.~F{\"u}rbass, P.~Ossenblok, M.~Hartmann, H.~Perko, A.~Skupch, G.~Lindinger,
  L.~Elezi, E.~Pataraia, A.~Colon, C.~Baumgartner \emph{et~al.}, ``Prospective
  multi-center study of an automatic online seizure detection system for
  epilepsy monitoring units,'' \emph{Clinical Neurophysiology}, vol. 126,
  no.~6, pp. 1124--1131, 2015.

\bibitem{alickovic2018performance}
E.~Alickovic, J.~Kevric, and A.~Subasi, ``Performance evaluation of empirical
  mode decomposition, discrete wavelet transform, and wavelet packed
  decomposition for automated epileptic seizure detection and prediction,''
  \emph{Biomedical signal processing and control}, vol.~39, pp. 94--102, 2018.

\bibitem{rastegari2016xnor}
M.~Rastegari, V.~Ordonez, J.~Redmon, and A.~Farhadi, ``Xnor-net: Imagenet
  classification using binary convolutional neural networks,'' in
  \emph{European Conference on Computer Vision}.\hskip 1em plus 0.5em minus
  0.4em\relax Springer, 2016, pp. 525--542.

\bibitem{lin2017towards}
X.~Lin, C.~Zhao, and W.~Pan, ``Towards accurate binary convolutional neural
  network,'' in \emph{Advances in Neural Information Processing Systems}, 2017,
  pp. 345--353.

\bibitem{yu2016binary}
S.~Yu, Z.~Li, P.-Y. Chen, H.~Wu, B.~Gao, D.~Wang, W.~Wu, and H.~Qian, ``Binary
  neural network with 16 mb rram macro chip for classification and online
  training,'' in \emph{2016 IEEE International Electron Devices Meeting
  (IEDM)}.\hskip 1em plus 0.5em minus 0.4em\relax IEEE, 2016, pp. 16--2.

\bibitem{courbariaux2016binarized}
M.~Courbariaux, I.~Hubara, D.~Soudry, R.~El-Yaniv, and Y.~Bengio, ``Binarized
  neural networks: Training deep neural networks with weights and activations
  constrained to+ 1 or-1,'' \emph{arXiv preprint arXiv:1602.02830}, 2016.

\bibitem{nakahara2018lightweight}
H.~Nakahara, H.~Yonekawa, T.~Fujii, and S.~Sato, ``A lightweight yolov2: A
  binarized cnn with a parallel support vector regression for an fpga,'' in
  \emph{Proceedings of the 2018 ACM/SIGDA International Symposium on
  Field-Programmable Gate Arrays}.\hskip 1em plus 0.5em minus 0.4em\relax ACM,
  2018, pp. 31--40.

\bibitem{ioffe2015batch}
S.~Ioffe and C.~Szegedy, ``Batch normalization: Accelerating deep network
  training by reducing internal covariate shift,'' \emph{arXiv preprint
  arXiv:1502.03167}, 2015.

\bibitem{abadi2016tensorflow}
M.~Abadi, P.~Barham, J.~Chen, Z.~Chen, A.~Davis, J.~Dean, M.~Devin,
  S.~Ghemawat, G.~Irving, M.~Isard \emph{et~al.}, ``Tensorflow: A system for
  large-scale machine learning,'' in \emph{12th $\{$USENIX$\}$ Symposium on
  Operating Systems Design and Implementation ($\{$OSDI$\}$ 16)}, 2016, pp.
  265--283.

\bibitem{cecotti2010convolutional}
H.~Cecotti and A.~Graser, ``Convolutional neural networks for p300 detection
  with application to brain-computer interfaces,'' \emph{IEEE transactions on
  pattern analysis and machine intelligence}, vol.~33, no.~3, pp. 433--445,
  2010.

\bibitem{khan2017focal}
H.~Khan, L.~Marcuse, M.~Fields, K.~Swann, and B.~Yener, ``Focal onset seizure
  prediction using convolutional networks,'' \emph{IEEE Transactions on
  Biomedical Engineering}, vol.~65, no.~9, pp. 2109--2118, 2017.

\end{thebibliography}
}
\end{document}